\documentclass[11pt]{article}
\usepackage{moriond,epsfig}

\newcommand{\eqref}[1]{Eq.~(\ref{#1})}

\newcommand{\be}{\begin{equation}}
\newcommand{\ee}{\end{equation}}
\newcommand{\ben}{\begin{equation*}}
\newcommand{\een}{\end{equation*}}
\newcommand{\bea}{\begin{eqnarray}}
\newcommand{\eea}{\end{eqnarray}}
\newcommand{\bean}{\begin{eqnarray*}}
\newcommand{\eean}{\end{eqnarray*}}
\newcommand{\brr}{\begin{array}}
\newcommand{\err}{\end{array}}
\newcommand{\bc}{\begin{center}}
\newcommand{\ec}{\end{center}}

\newcommand{\ie}{\mbox{\it i.e.~}}
\newcommand{\cf}{\mbox{\it cf.~}}
\newcommand{\vev}[1]{\mbox{$\langle #1 \rangle $}}

\newcommand{\bk}{{\mathbf k}}

\newcommand{\HH}{{\mathcal H}}

\newcommand{\de}{\delta}

\newcommand{\Om}{\Omega}

\begin{document}
\vspace*{4cm}
\title{GRAVITATIONAL WAVES FROM COSMOLOGICAL PHASE TRANSITIONS}

\author{C. CAPRINI}

\address{CEA, IPhT \& CNRS, URA 2306, F-91191 Gif-sur-Yvette, France}

\maketitle\abstracts{First order phase transitions in the early universe can give rise to a stochastic background of gravitational waves. A hypothetical first order electroweak phase transition is particularly interesting in this respect, since the signal is in the good frequency range to be detectable by the space interferometer LISA. Three main processes lead to the production of the gravitational wave signal: the collision of the broken phase bubbles, the magnetohydrodynamical turbulence in the plasma stirred by the bubble collisions, and the magnetic fields amplified by the magnetohydrodynamical turbulence. The main features of the gravitational wave spectrum, such as the peak frequency, the amplitude, and the slopes both at low and high wave-number can be predicted by general arguments based on the characteristics of the source: in particular, the structure of its space and time correlation. We find that the gravitational wave signal from a first order phase transition occurring at electroweak symmetry breaking falls into the LISA sensitivity range if the phase transition lasts for about one hundredth of the Hubble time and the energy density of the turbulent motions is about twenty percent of the total energy density in the universe at the phase transition time.}

\section{Introduction}

It is likely that in the years to come the interferometers LIGO \cite{LIGO} and VIRGO \cite{VIRGO} will provide the first direct detection of gravitational waves (GWs). Terrestrial interferometers operate in the frequency range 10-1000 Hz, while the space-based interferometer LISA \cite{LISA} will have its best sensitivity around the milliHertz frequency. A possible target of these experiments is a stochastic background of GWs of cosmological origin \cite{maggiore}. The detection of such a background, relic of the early universe, would have a profound impact both on cosmology and on high energy physics. Once GWs have been emitted, in fact, they propagate freely through the universe, carrying direct information on the physical process that generated them: their detection would therefore provide a new way of probing the primordial universe, and correspondingly physics at very high energies, which would not be accessible otherwise. 

One of the processes that can generate such a stochastic background of GWs is a relativistic first order phase transition \cite{hogan83,witten,hogan86,TW,beta}. In the course of its adiabatic expansion, the universe might have undergone several phase transitions driven by the temperature decrease. The nature of the phase transitions depends on the particle theory model, but if they are first order they proceed through the nucleation of broken phase bubbles, which is a very violent and inhomogeneous process capable of sourcing GWs \cite{Kos92,Kos93,Kam94}. In the following we review the different mechanisms of generation of GWs by a first order phase transition, and show that the characteristic frequency and amplitude of the GW signal are related respectively to the temperature at which the phase transition occurs and to its strength. We also review the analytical method developed in Refs.~\cite{bubble,thomas,MHD}, which predicts the main features of the GW spectrum starting from a description of the source which we tried to make as simple and as model independent as possible. The GW signal depends explicitly on a few free parameters: under specific choices for these parameters, it can be potentially interesting for observation with LISA \cite{apreda,nicolis,kahn1} or, more speculatively, advanced LIGO. 

\section{Gravitational wave background of cosmological origin}

In the cosmological context GWs are small perturbations of the FRW metric represented by the transverse-traceless tensor $h_{ij}$ which is first order in cosmological perturbation theory:
\be
ds^2=a^2(t)(dt^2-(\delta_{ij}+2h_{ij})dx^idx^j)\,,~~~~~~~~~~{h_i}^i={{h_i}^j}_{|j}=0\,,
\ee
where we assume flat spatial sections and $t$ denotes conformal time. Inserting the perturbed metric into Einstein's equations $G_{\mu\nu}=8\pi G\, T_{\mu\nu}$ one obtains the evolution equation for the tensor perturbation, which in a radiation dominated universe takes the form
\be
\ddot{h}_{ij}(\bk,t)+\frac{2}{t}\dot{h}_{ij}(\bk,t)+k^2h_{ij}(\bk,t)= 8\pi G a^2 T_{ij}^{(TT)}(\bk,t)\,,
\label{gweq}
\ee
where $T_{ij}^{(TT)}(\bk,t)$ is the transverse-traceless part of the energy momentum tensor sourcing the GWs, \ie the tensor anisotropic stress. A source of GWs operating in the primordial universe is described as a stochastic process, and generates a stochastic background of GWs statistically homogeneous, isotropic, unpolarised and assumed to be gaussian. The energy density of the GWs, normalised to the critical energy density of the universe today $\rho_c$, is given by the integral over wave-number of the power spectrum 
\bea
& &\Omega_{\rm GW}=\frac{\langle\dot h_{ij}\dot h^{ij}\rangle}{8\pi G\rho_c a^2}=\int\frac{{\rm d}k}{k}\frac{{\rm d}\Omega_{\rm GW}}{{\rm d}\log k}\,,~~~~~{\rm where}~~~~~\frac{{\rm d}\Omega_{\rm GW}}{{\rm d}\log k}=\frac{k^3|\dot h|^2}{2(2\pi)^3G\rho_c a^2}\,,\label{spectrum1} \\
& &{\rm and}~~~~~~~~~~ \langle \dot{h}_{ij}(\mathbf{k},t)\dot{h}_{ij}^*(\mathbf{q},t)\rangle=(2\pi)^3\delta(\mathbf{k}-\mathbf{q})|\dot{h}(k,t)|^2\,. \label{spectrum2}
\eea
If we consider a source of GWs active at a given time $t_*$ while the universe is in a phase of standard FRW expansion (\ie anytime besides inflation), then causality constrains the characteristic frequency of the emitted GWs to be larger than the causal horizon of the universe at the time of generation:  $k_*\geq \HH_*$, where $\HH_*$ denotes the conformal Hubble parameter and $k_*$ is the comoving wave-number. For example, GWs generated at the electroweak phase transition (EWPT) at $T_*\simeq 100$ GeV must have a characteristic frequency $k_*\geq 10^{-5}$Hz, while the characteristic frequency of GWs generated at the QCD phase transition at $T_*\simeq 100$ MeV can be much lower, $k_*\geq 10^{-8}$Hz. This estimation, based on the causality argument, shows that GWs generated during the EWPT are potentially interesting for detection with the space interferometer LISA, which operates in the frequency window from $10^{-4}$ to 1 Hz. On the other hand, GW production at the QCDPT can fall into the frequency range of detection with pulsar timing array \cite{PTA}. We now proceed to analyse which kind of processes can act as sources of GWs during a relativistic phase transition occurring in the early universe.

\section{Gravitational waves from phase transitions}
\label{sec:GWPT}

There are a variety of processes related to primordial phase transitions that can lead to the production of GWs, as for example cosmic strings \cite{strings} or scalar field relaxation \cite{relax}. Here we concentrate specifically on the processes driven by bubble nucleation during a first order phase transition. The EWPT in the standard model is a crossover, and it is not expected to lead to any appreciable cosmological signal; however, deviations from the standard model in the Higgs sector or introducing supersymmetry can lead to a first order EWPT \cite{deviations}. Similarly, the QCDPT which is also predicted to be a crossover by lattice simulations \cite{lattice}, can become first order if the neutrino chemical potential is sufficiently large \cite{schwarz}. GWs detection would help to probe the nature of these phase transitions, and provide interesting information on the underlying particle theory.

A first order phase transition proceeds through the nucleation of true-vacuum bubbles, which in the end of the transition collide and convert the entire universe to the broken phase. The collisions break the spherical symmetry of the bubble walls, generating a non-zero anisotropic stress which acts as a source of GWs \cite{Kos92,Kos93,Kam94,bubble}. Moreover, bubble collision causes an injection of energy in the primordial plasma, which has a very high Reynolds number (of the order of $10^{13}$ at 100 GeV and at the typical scale of the bubbles \cite{MHD}): the energy injection leads to the formation of magnetohydrodynamic (MHD) turbulence in the fluid, which also sources GWs through the anisotropic stresses of the chaotic fluid motions \cite{hogan832,Kos02,dolgov,MHDold,gogob,THelical,THelical2,MHD}. MHD turbulence also leads to the amplification of small magnetic fields generated by charge separation at the bubble wall, which have non-zero anisotropic stress and act as a source of GWs \cite{hogan832,Kos02,dolgov,MHDold,gogob,THelical,THelical2,MHD}. 

There are in summary three processes which can lead to the production of GWs towards the end of a first order phase transition. They are all related to the collision of bubbles, which involves two quantities: the duration of the phase transition, commonly denoted by the parameter $\beta^{-1}$, and the typical size of the bubbles at the moment of collision, $R_* \simeq v_b \,\beta^{-1}$, where $v_b$ is the bubble wall velocity. The characteristic frequency of the GWs generated by the three processes can correspond either to the duration of the phase transition or to the bubble size: $k_*\simeq \beta\,,$ $R_*^{-1}$, depending on the details of the time evolution of the source (\cf section \ref{sec:GWspec}). Assuming for the moment $k_*\simeq \beta$, one obtains the following order of magnitude estimate of the characteristic frequency in Hz:
\be
k_*\simeq 10^{-2}\,\frac{\beta}{\mathcal{H}_*}\,\frac{T_*}{100\,{\rm GeV}}\,{\rm mHz}\,.
\label{kstar}
\ee
The parameter $\beta/\HH_*$ represents the ratio of the duration of the phase transition with respect to the Hubble time. Since the entire universe must  be converted to the broken phase, the phase transition must complete faster than Hubble time: a typical value for the EWPT is\cite{beta} $\beta/\HH_*\simeq 100$. From Eq.~(\ref{kstar}) one gets that the characteristic frequency of GWs emitted at the EWPT is of the order of the milliHertz, and falls in the frequency range of the space interferometer LISA, which covers the frequency region $10^{-4}\mathrm{Hz}\leq k\leq 1 \mathrm{Hz}$. LISA reaches its best sensitivity precisely around a milliHertz, where it can detect GWs with amplitude corresponding to $\Omega_{\rm GW}\simeq 10^{-12}$. 

Starting from the perturbed Einstein equations $\de G_{\mu\nu}=8\pi G\, \de T_{\mu\nu}$, one can give a simple order of magnitude estimate of the GW amplitude, which shows how the result depends on the duration and the energy density of the GW source. Since we are merely interested in determining the scaling of the GW amplitude, we rewrite Einstein equations simply as $\beta^2 h \sim 8\pi G \,T$, where $h$ denotes the amplitude of the tensor perturbation, $T$ the energy momentum tensor, and we inserted $1/\beta$ as the characteristic time on which the perturbation is evolving (we have dropped indices for simplicity). This suggests that $\dot h \sim 8\pi G \,T/\beta$, and the GW energy density becomes then $\rho_{\rm GW}\sim \dot h^2/(8\pi G)\sim 8 \pi G\, T^2/ \beta^2$. For the three processes under analysis here, the energy momentum tensor of the GW source can be rewritten in all generality as $T/ \rho_* \sim \Omega_s^*$, where $\rho_*$ denotes the energy density in the universe at the GW generation time, and the parameter $\Omega_s^*$ denotes the relative energy density available in the source for the GW generation. We can now rearrange the equation for $\rho_{\rm GW}$ accounting for the fact that the phase transition takes place in the radiation-dominated universe, that the GW energy density also evolves like radiation, and inserting Friedmann equation $3\HH_*=8\pi G \rho_*$. It becomes then
\be
\Om_{\rm GW} \sim \Om_{\rm rad} \left(\frac{\HH_*}{\beta}\right)^2 (\Om_s^*)^2\,,
\label{Omgw}
\ee
where $\Om_{\rm rad}$ denotes the radiation energy density parameter today. The above equation shows that the GW energy density scales like the square of the ratio of the GW source duration and the Hubble time, and the square of the energy density in the source. Using $h^2\Om_{\rm rad}=4.2\cdot 10^{-5}$ and $\HH_*/\beta\simeq 0.01$, it turns out that a GW source with relative energy density of the order of $\Om_s^*\sim 0.1$ could generate a GW signal above the lowest sensitivity of the space interferometer LISA. A detectable signal can therefore arise only from very energetic processes, which must involve at least 10\% of the total energy density in the universe: namely, the phase transition must be strongly first order. 

\section{The gravitational wave power spectrum} 
\label{sec:GWspec}

Having demonstrated that GWs generated by a first order phase transition have a characteristic frequency and amplitude which can be interesting for detection, we now proceed to evaluate the main features of the GW power spectrum. The aim of this section is to show how the form of the power spectrum depends on the characteristics of the stochastic source, in particular the structure of its spatial and temporal correlation. It is based on the work done in Refs.~\cite{bubble,thomas,MHD}.

As derived in \cite{MHD}, Eq.~(\ref{gweq}) can be solved in the radiation era (neglecting changes in the number of effective relativistic degrees of freedom), and once combined with definitions (\ref{spectrum1}) and (\ref{spectrum2}) it leads to the following formula for the GW power spectrum today \cite{}:
\be
\frac{{\rm d}\Omega_{\rm GW}}{{\rm d} \log k} =\frac{4\Omega_{\rm rad} }{3 \pi^2}  \ k^3 \  \int_{t_{\rm in}}^{t_{\rm fin}} \frac{dt_1}{t_1} \int_{t_{\rm in}}^{t_{\rm fin}} 
\frac{dt_2}{t_2}\cos[k(t_2-t_1)] \Pi(k,t_1,t_2) \,,
\label{Omspectrum}
\ee
where $\Pi(k,t_1,t_2)$ is the unequal time power spectrum of the source anisotropic stress,
\be
\langle \Pi_{ij}({\bf{k}},t_1)\Pi_{ij}^*({\bf{q}},t_2) 
 \rangle= (2\pi)^3 \delta({\bf{k}}-{\bf{q}}) \Pi(k,t_1,t_2) \, ,
\label{Pispectrum}
\ee
and we denote $\Pi_{ij}$ the dimensionless anisotropic stress: $T^{(TT)}_{ij}=(\rho +p)\,\Pi_{ij}$ ($\rho$ and $p$ are the energy density and pressure of the primordial relativistic fluid, \cf Eq.~(\ref{gweq}) and Ref.~\cite{MHD}). The transverse traceless part of the spatial energy momentum tensor of the source contributes to the tensor anisotropic stress. In the bubble collisions case the spatial energy momentum tensor is given by $T_{ij}=(\rho+p)\,v_i\,v_j/(1-v^2)$, where $v_i$ is the velocity of the relativistic fluid at the bubble wall position \cite{bubble}. In the case of MHD turbulence, the energy momentum tensor can be decomposed into a part representing the turbulent velocity field and a part representing the magnetic field (for details, see \cite{MHD}). The turbulent velocity field part takes the form $T_{ij}=(\rho+p)\,v_iv_j$ (where we omit the gamma factor since we assume only mildly relativistic velocities), and the magnetic field one takes the form $T_{ij}=B_iB_j/(4\pi)$. We see that for all the three GW generation processes the tensor anisotropic stress is radiation-like, since the phase transition takes place in the radiation dominated universe. The parameter $\Omega_s^*$ introduced in section \ref{sec:GWPT} is therefore given simply by $\Omega_s^*\sim \vev{v^2}$ in the case of the turbulence and $\Omega_s^*\sim \vev{b^2}$ in the case of the MHD processed magnetic field, where $b_i=B_i/\sqrt{\rho_{\rm rad}}$ is the dimensionless magnetic field parameter \cite{MHD}. In the bubble case, the definition of $\Omega_s^*$ is a bit more involved, depending on whether the phase transition happens in a thermal bath or in vacuum: it is anyway related to the ratio of the kinetic energy density due to the bubble wall motions and the total energy density in the universe \cite{thomas}.

Eq.~(\ref{Omspectrum}) shows that the GW power spectrum is given by the double integral of the Green function of Eq.~(\ref{gweq}) multiplied by the unequal time anisotropic stress power spectrum, and its general shape as a function of wave-number $k$ depends on the interplay among these two quantities. In particular, the spatial and temporal correlation of the source, together with its overall time evolution, determine the $k$ and time dependence of the anisotropic stress power spectrum both at equal and unequal time, and in turns the GW spectrum. Much of the analytical work of Refs.~\cite{bubble,thomas,MHD} has dealt with the problem of modeling the statistical source for bubble collisions and MHD turbulence, and here we summarise the results obtained there.

\begin{figure}
\psfig{figure=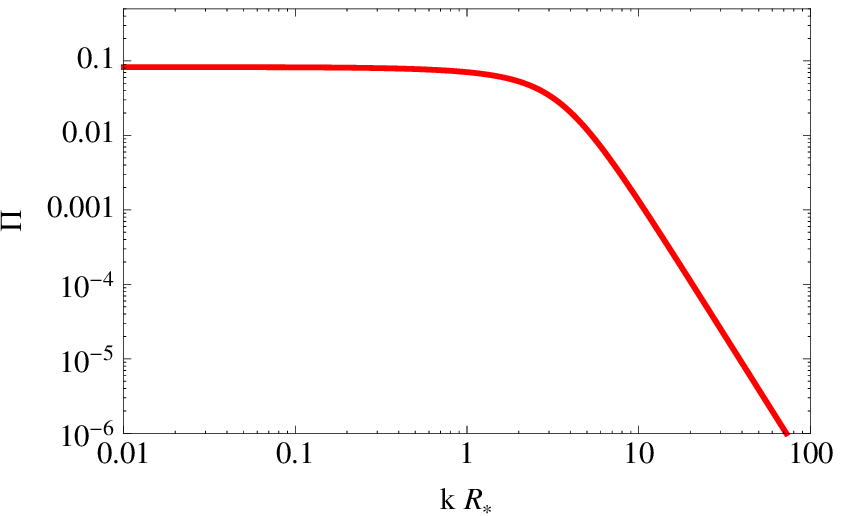,height=1.9in}\hspace*{0.5cm}
\psfig{figure=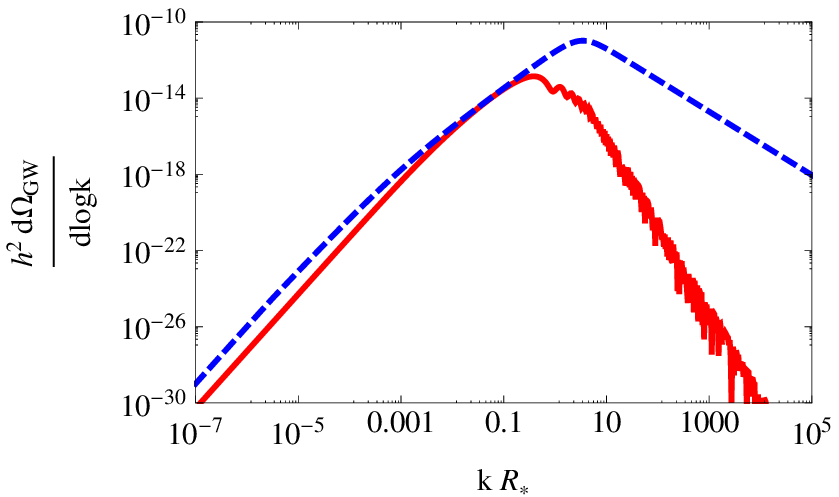,height=1.9in}
\caption{Left panel: the anisotropic stress power spectrum at equal time as a function of the dimensionless variable $k R_*$ for the MHD turbulence. On scales larger than the characteristic scale $R_*$ the spectrum is flat since the source is uncorrelated, while on scales smaller than $R_*$ the spectrum decays as $k^{-11/3}$ (we assumed Kolmogorov turbulence). Right panel: the qualitative behaviour of the GW spectrum from the MHD turbulent source modeled as completely coherent (red, solid line) and as top-hat decorrelation (blue, dashed line). The peak position corresponds to the characteristic time-scale of the source $k_*\simeq \beta$ for the coherent case, and to the characteristic length-scale $k_*\simeq 1/R_*$ for the top-hat decorrelation case. The low frequency increase is $k^3$ by causality, and the high frequency slope depends both on the time decorrelation assumption and on the $k$-dependence of the source power spectrum (left panel). 
\label{fig:Pi}}
\end{figure}

First of all, the $k$-dependence of the equal time power spectrum $\Pi(k,t_1,t_1)$ is relatively easy to determine, following mainly from a causality argument. The bubble collisions and subsequent MHD turbulence are causal processes, characterised by a typical length-scale:  the size of the bubbles at the moment of collision $R_*$, which also corresponds to the scale of energy injection in the primordial fluid generating the MHD turbulence. On scales larger than $R_*$, the anisotropic stresses are not correlated and we expect the power spectrum to be flat, up to the wave-number corresponding to the inverse characteristic scale $R_*^{-1}$. Beyond $k\simeq R_*^{-1}$, the power spectrum decays with a slope that depends on the details of the source, and turns out to be $k^{-4}$ for bubble collisions \cite{bubble}, and $k^{-11/3}$ for MHD turbulence \cite{MHD}. The spatial correlation structure of the source completely determines the $k$-dependence of $\Pi(k,t_1,t_1)$, shown in Fig.~\ref{fig:Pi} for the MHD turbulence case. 

Concerning the time dependence of $\Pi(k,t_1,t_1)$, it is due to the overall evolution of the GW source in time. In the case of bubble collisions, the source turns off right at the end of the phase transition, and lasts therefore for much less than one Hubble time. The overall time dependence of the source in this case comes mainly from the bubble expansion \cite{thomas}. In the case of MHD turbulence on the other hand, the dissipation of the turbulent motions in the primordial fluid is not very efficient due to the extremely low viscosity of the fluid itself \cite{MHD}. MHD turbulence is therefore maintained in the primordial fluid for several Hubble times, and the overall time dependence of the source depends both on the slow decay of the turbulent motions and on the growth of the turbulence characteristic scale associated with the decay. The influence of the total duration of the source on the shape of the GW spectrum shows up mainly at very large scales: if the source lasts several Hubble times, the GW signal is amplified on scales larger than the Hubble scale at the phase transition time $k< \HH_*$ \cite{MHD}. 

Moreover, in order to find the GW spectrum one needs to know the unequal time power spectrum of the anisotropic stress, as given in Eq.~(\ref{Omspectrum}). This is much less obvious to determine, but one can identify a few simple models which are easy to deal with analytically and represent the main properties of the source. These have been discussed in \cite{bubble,thomas,MHD}, and in what follows we present the main results. In the bubble collision case, the source can be modeled as {\it completely coherent}, meaning that its time dependence is deterministic. In this case, the unequal time anisotropic stress power spectrum can be decomposed in terms of the equal time one as
\be
\Pi(k,t_1,t_2)=\sqrt{\Pi(k,t_1,t_1)}\sqrt{\Pi(k,t_2,t_2)}\,.
\label{Picoh}
\ee
This is a consequence of two main facts: first of all, the signal comes from the individual collision events, and each collision event can be assumed to be uncorrelated in time with the others; second, each collision event can also be assumed to be coherent, since the time evolution of the anisotropic stress related to the collisions is deterministic (it is basically only due to the growth in time of the bubbles \cite{thomas}). In the case of MHD turbulence, the situation is different. The turbulence can be viewed as a superposition of eddies of different size, each with its proper turnover time related to the eddy size. In the Kraichnan model \cite{kraichnan}, the eddy turnover time is the typical decorrelation time of turbulent motions. Therefore, the source has a finite decorrelation time which depends on the eddy size: this can be modeled with a {\it top-hat decorrelation}: 
\be
\Pi(k,t_1,t_2)=\{\Pi(k,t_1,t_1)\Theta[t_2-t_1]\Theta[1-k(t_2-t_1)]+t_1\leftrightarrow t_2\} \,,
\label{Pitop}
\ee
meaning that the source is correlated for time differences smaller than about one wavelength \cite{MHD}: $|t_1-t_2|<k^{-1}$.  

The behaviour of the anisotropic stress unequal time power spectrum strongly influences the GW spectrum, and in particular the position of the peak. For a coherent source, the GW spectrum becomes the square of the temporal Fourier transform of the source \cite{thomas}, as can be seen by inserting Eq.~(\ref{Picoh}) into (\ref{Omspectrum}). The source is characterised by the spatial correlation scale $R_*$ and the temporal correlation scale $\beta^{-1}$, related by $R_*= v_b \beta^{-1}$. Since $v_b\leq 1 $, one has $R_* < \beta^{-1}$. On scales larger than both the characteristic spatial correlation scale $R_*$ and temporal correlation scale $\beta^{-1}$, the space and time Fourier transforms of the source are constant because the source is not correlated (white noise). Therefore, for wave-numbers $k<  \beta< 1/R_*$, the GW spectrum simply increases as $k^3$ (\cf Eq.~(\ref{Omspectrum})). However, for $k> \beta$ the time Fourier transform is no longer constant and starts to decay as a power law, the exponent depending on the time differentiability properties of the source \cite{thomas}. In the bubble collision case, it turns out that this implies a $k^{-1}$ decay for the GW spectrum at intermediate scales $\beta<k<1/ R_*$: the peak of the GW spectrum corresponds therefore to the characteristic time of the source \cite{thomas}, $k_*\simeq \beta$. For a source with finite decorrelation time as in Eq.~(\ref{Pitop}), on the other hand, the GW spectrum is no longer related to the temporal Fourier transform of the source, and the situation changes. The spectrum still increases as $k^3$ on large scales, but the characteristic time of the source $\beta^{-1}$ does not lead to any feature in the spectrum, which peaks instead at the inverse characteristic scale of the source $k_*\simeq R_*^{-1}$. The qualitative shape of the GW spectra coming from a coherent source and a top-hat decorrelating one is shown in Fig.~\ref{fig:Pi}. 

\section{Results}

\begin{figure}
\psfig{figure=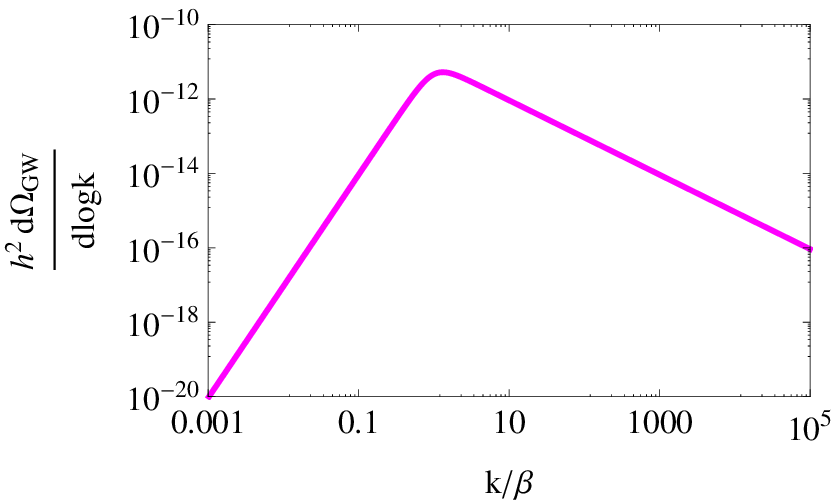,height=1.9in}\hspace*{0.5cm}
\psfig{figure=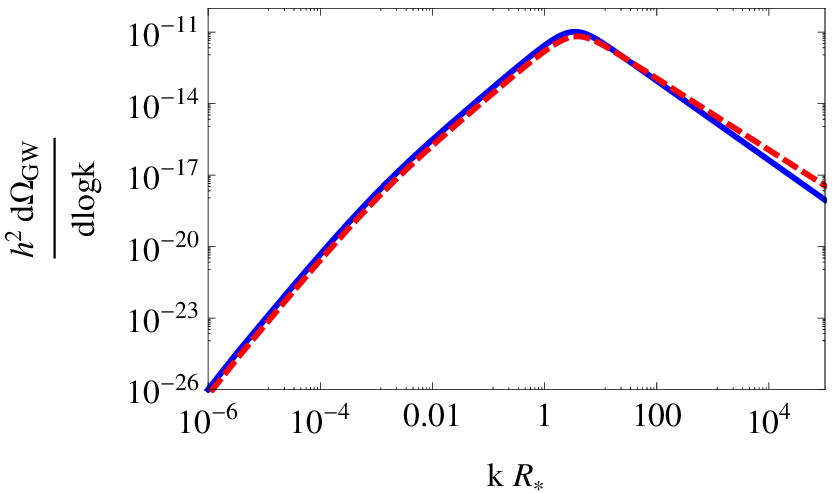,height=1.9in}
\caption{Left panel: the GW power spectrum generated by bubble collisions as a function of the dimensionless variable $k/\beta$. The spectrum increases as $k^3$ at low frequency, peaks at $k_*\simeq \beta$ since the source is coherent, and decreases at high frequency as $k^{-1}$, due both to the fact that the source is coherent and to the thin wall approximation. Right panel: the GW power spectrum generated by MHD turbulence as a function of $k R_*$: blue, solid: the turbulent velocity field and red, dashed: the magnetic field. The spectra increase as $k^{3}$ at low frequency, turn to a $k^2$ increase at subhorizon scales $k> \HH_*$ because the source lasts for several Hubble times, and peak at a frequency corresponding to the characteristic scale of the source $k_*\simeq 2\pi/R_*$ due to the top-hat decorrelation holding in the case of MHD turbulence. At high frequency, the spectra decay as $k^{-5/3}$ for the kinetic turbulence (Kolmogorov spectrum for the source) and as $k^{-3/2}$ for the magnetic field (Iroshnikov-Kraichnan spectrum for the source). The duration of the phase transition and the energy density of the sources are given the values $\HH_*/\beta=0.01$ and $\Omega_s^*=0.2$.
\label{fig:spectra}}
\end{figure}

Fig.~\ref{fig:spectra} shows the GW spectra generated by bubble collisions and MHD turbulence, as derived in Refs.~\cite{bubble,thomas,MHD}. The parameter representing the duration of the phase transition is set to $\HH_*/\beta=0.01$, and the parameter representing the relative energy density available in the source for the GW generation is set to $\Omega_s^*=0.2$ in both cases (\cf Eq.~\ref{Omgw}). This high value of $\Omega_s^*$ implies a strongly first order phase transition, for which the vacuum energy density is about one third of the radiation one $\alpha=1/3$, and the bubble wall speed is close to the speed of light, $v_b=0.87$. Correspondingly, the mean velocity of the turbulent motions must be of the order of the speed of sound, $\vev{v^2}=1/3$, and equipartition is assumed among the kinetic and the magnetic energies in the turbulence such that $\vev{b^2}\simeq \vev{v^2}$. 

In the two GW spectra of Fig.~\ref{fig:spectra} we can distinguish the features presented in the previous sections. Both spectra rise as $k^3$ for small wave-numbers: this is simply the phase space volume (\cf Eq.~(\ref{Omspectrum})) combined with the $k$-dependence of the anisotropic stress power spectrum, which is flat at small wave-numbers because of causality. In the GW spectrum generated by bubble collisions the $k^3$ slope is maintained up to the peak, while in the one generated by MHD turbulence the slope changes to $k^2$ at sub-horizon scales $k>\HH_*$: this difference is due to the fact that bubble collisions are a short lasting source, while MHD turbulence acts a source of GWs for several Hubble times \cite{MHD}. The GW spectrum due to bubble collisions peaks at a wave-number corresponding to the duration of the source $k_*\simeq \beta$, since the source is modeled as completely coherent in its time decorrelation; while the GW spectrum due to MHD turbulence peaks at a wave-number corresponding to the characteristic length-scale of the source, given by the bubble size at the end of the phase transition: $k_*\simeq 2\pi/R_*$, since the source is modeled following the top-hat Ansatz and has a finite decorrelation time corresponding to the eddy turnover time \cite{MHD}. At frequencies smaller than the peak, the GW spectrum from bubble collisions decays as $k^{-1}$: this decay is related both to the fact that the source is coherent and to the thin wall approximation, which has been inserted in the analytical analysis to recover the result of numerical simulations \cite{huber,thomas}. The GW spectrum from MHD turbulence decays at high frequencies with slopes that also depend both on the top-hat decorrelation structure and on the source power spectrum: they turn out to be $k^{-5/3}$ for the kinetic turbulence (coming directly from the assumption of a Kolmogorov spectrum) and $k^{-3/2}$ for the magnetic field (due to the assumption of an Iroshnikov-Kraichnan spectrum) \cite{MHD}.

\begin{figure}
\begin{center}
\psfig{figure=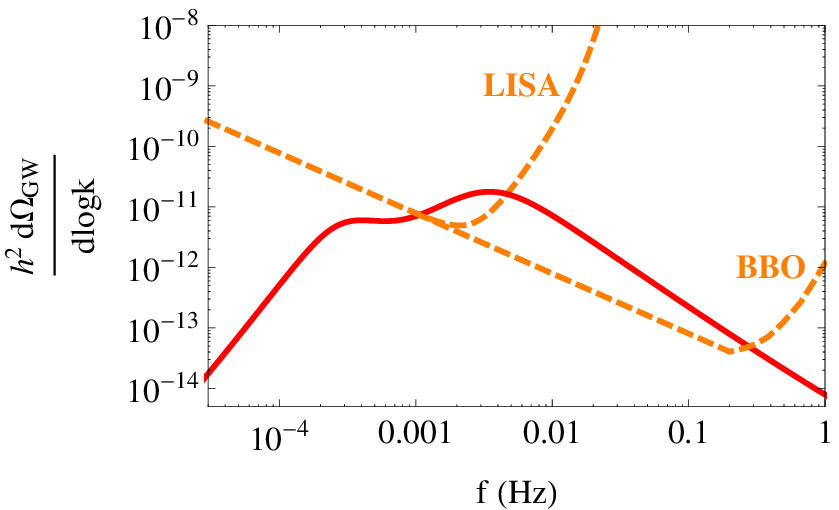,height=2in}
\psfig{figure=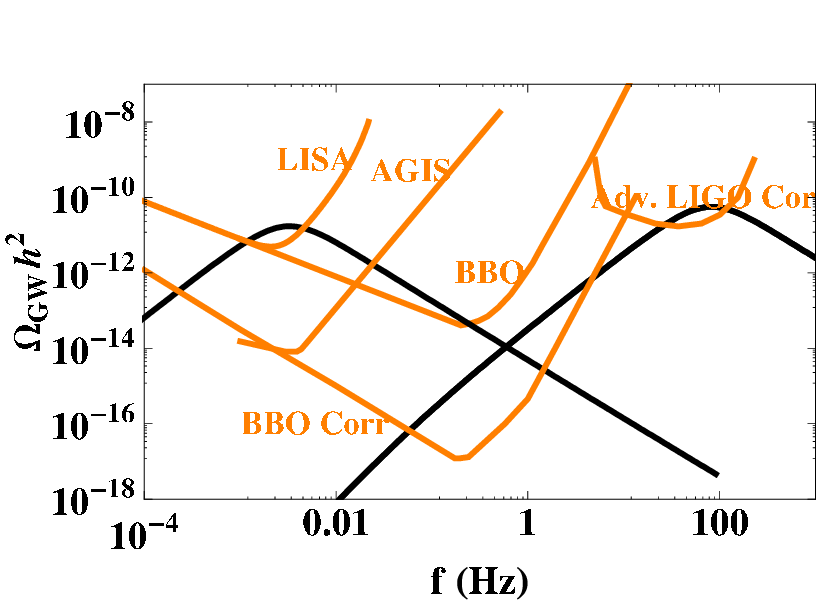,height=1.7in}
\caption{Left panel: the total GW spectrum for bubble collisions and MHD turbulence generated at a strongly first order EWPT as a function of frequency in Hz. The duration of the phase transition and the energy density of the sources are given the values $\HH_*/\beta=0.01$ and $\Omega_s^*=0.2$, and the signal falls into the sensitivity range of the space interferometer LISA. Right panel: the total GW spectrum from MHD turbulence only, generated at a strongly first order EWPT (left black curve) and at a hypothetical first order phase transition at temperature $T_*= 5\cdot 10^6$ GeV. The parameters are $\HH_*/\beta=0.01$ and $\Omega_s^*=0.2$ for the EWPT and $\HH_*/\beta=0.02$ and $\Omega_s^*=0.2$ for the PT at $T_*= 5\cdot 10^6$ GeV. Figure from$^7$.
\label{fig:final}}
\end{center}
\end{figure}

Fig.~\ref{fig:final} shows the final GW power spectrum given by the sum of the bubble collisions signal and the MHD turbulence signal for the EWPT at $T_*\simeq 100$ GeV. The parameters are again $\HH_*/\beta=0.01$ and $\Omega_s^*=0.2$, so that the phase transition is assumed to be strongly first order. The signal is compared with the sensitivity curves of LISA and BBO: for a strongly first order EWPT the signal falls into the sensitivity range of LISA. Clearly this can only be achieved in the context of theories that go beyond the standard model of particle physics. Fig.~\ref{fig:final} also shows an even more speculative case of a phase transition occurring at temperature $T_*= 5\cdot 10^6$ GeV, with $\HH_*/\beta=0.02$ and $\Omega_s^*=0.2$: in this case, the signal could be interesting for advanced LIGO \cite{chialva}.

%\section*{Acknowledgments}


\section*{References}

\begin{thebibliography}{99}

\bibitem{LIGO} URL http://www.ligo.caltech.edu/
\bibitem{VIRGO} URL http://www.virgo.infn.it/
\bibitem{LISA} URL http://lisa.jpl.nasa.gov/
\bibitem{maggiore} For a review, see M. Maggiore, {\it Phys. Rept.} {\bf 331}, 283 (2000) and references therein.

\bibitem{hogan83} C.J. Hogan, {\it Phys. Lett.} B {\bf 133}, 172 (1983). 
\bibitem{witten} E. Witten, {\it Phys. Rev.}  D {\bf 30}, 272 (1984).
\bibitem{hogan86} C.J. Hogan, {\it MNRAS} {\bf 218}, 629 (1986). 
\bibitem{TW} M.S. Turner and F. Wilczek, {\it Phys. Rev. Lett.} {\bf 65}, 3080 (1990).
\bibitem{beta} M.S. Turner {\it et al, Phys. Rev.} D {\bf 46}, 2384 (1992)
\bibitem{Kos92} A. Kosowsky {\it et al, Phys. Rev.} D {\bf 45}, 4514 (1992).
\bibitem{Kos93} A. Kosowsky and M. S. Turner, {\it Phys. Rev.} D {\bf 47}, 4372 (1993). 
\bibitem{Kam94} M. Kamionkowski {\it et al, Phys. Rev.} D {\bf 49}, 2837 (1994).  


\bibitem{bubble} C. Caprini {\it et al}, {\it Phys. Rev.}  D {\bf 77}, 124015 (2008).
\bibitem{thomas} C. Caprini {\it et al}, {\it Phys. Rev.} D {\bf 79}, 083519 (2009).
\bibitem{MHD} C. Caprini {\it et al}, {\it JCAP}, {\bf 0912} 024 (2009).

\bibitem{apreda} R. Apreda et al, {\it Class. Quant. Grav.} {\bf 18}, L155 (2001). 
\bibitem{nicolis} A. Nicolis, {\it Class. Quant. Grav.} {\bf 21}, L27 (2004). 
\bibitem{kahn1} T. Kahniashvili {\it et al, Phys. Rev.} D {\bf 78}, 043003 (2008).

\bibitem{PTA} G. Hobbs {\it et al}, {\it Class. Quantum Grav.} {\bf 27}, 084013 (2010).
\bibitem{strings} See LIGO Scientific Collaboration: B. Abbott {\it et al}, {\it  Phys. Rev.} D {\bf 80}, 062002 (2009) and references therein.
\bibitem{relax} L.M. Krauss, {\it Phys. Lett.} B {\bf 284}, 229 (1992); E. Fenu {\it et al}, {\it JCAP} {\bf 10}, 005 (2009).
\bibitem{deviations} See for example C. Grojean {\it et al, Phys. Rev.} D {\bf 71}, 036001 (2005); S. J. Huber and T. Konstandin, {\it JCAP} {\bf 0805}, 017 (2008).
\bibitem{lattice} See for example Y. Aoki {\it et al}, {\it JHEP} {\bf 0906}, 088 (2009).
\bibitem{schwarz} D. J. Schwarz and M. Stuke, {\it JCAP} {\bf 11}, 025 (2009).

\bibitem{hogan832} C.J. Hogan, {\it Phys. Rev. Lett.}  {\bf 51}, 1488 (1983). 
\bibitem{Kos02} A. Kosowsky {\it et al, Phys. Rev.} D {\bf 66}, 024030 (2002).
\bibitem{dolgov} A.D. Dolgov {\it et al, Phys. Rev.} D {\bf 66}, 103505 (2002).
\bibitem{MHDold} C. Caprini and R. Durrer, {\it Phys. Rev.} D {\bf 74}, 063521 (2006).
\bibitem{gogob} G. Gogoberizde {\it et al, Phys. Rev.} D {\bf 76}, 083002 (2007).
\bibitem{THelical} T. Kahniashvili {\it et al, Phys. Rev. Lett.} {\bf 100}, 0231301 (2008). 
\bibitem{THelical2} T. Kahniashvili {\it et al, Phys. Rev.} D {\bf 78}, 123006 (2008). 

\bibitem{kraichnan} R.H. Kraichnan, {\it Phys. Fluid} {\bf 7}, 1163 (1964).
\bibitem{huber} S. J. Huber and T. Konstandin, {\it JCAP} {\bf 0809}, 022 (2008).

\bibitem{chialva} for even higher temperature, see for example D. Chialva, arXiv:1004.2051.

\end{thebibliography}
\end{document}